# Adaptive Reversible Watermarking for Medical Videos Based on Linear Prediction

Hamidreza Zarrabi, Ali Emami, Nader Karimi, Shadrokh Samavi
Department of Electrical and Computer Engineering, Isfahan University of Technology, Isfahan, Iran

*Abstract*— **Reversible video watermarking can guarantee that the watermark logo and the original frame can be recovered from the watermarked frame without any distortion. Although reversible video watermarking has successfully been applied in multimedia, its application has not been extensively explored in medical videos. Reversible watermarking in medical videos is still a challenging problem. The existing reversible video watermarking algorithms, which are based on error prediction expansion, use motion vectors for prediction. In this study, we propose an adaptive reversible watermarking method for medical videos. We suggest using temporal correlations for improving the prediction accuracy. Hence, two temporal neighbor pixels in upcoming frames are used alongside the four spatial rhombus neighboring pixels to minimize the prediction error. To the best of our knowledge, this is the first time this method is applied to medical videos. The method helps to protect patients' personal and medical information by watermarking, i.e., increase the security of Health Information Systems (HIS). Experimental results demonstrate the high quality of the proposed watermarking method based on PSNR metric and a large capacity for data hiding in medical videos.**

## I. INTRODUCTION

Watermarking in a video destroys the video due to embedding process. This distortion is not desired in medical applications, because small distortion in the video may have an impact on the decision of physician. One proposed solution is the use of reversible watermarking, such that both the original video and watermark can be restored from the watermarked video. Therefore, reversible video watermarking is inevitable in health information system (HIS) which has attracted a lot of attention in the research community.

Many reversible image watermarking methods exist in the literature. In [1], the watermark is embedded in integer wavelet transform (IWT) coefficient, and genetic programming as learning capability is exploited for intelligent selection of IWT coefficients. In [2], the initial region of interest and non-region of interest is automatically extracted with adaptive threshold detector algorithm. Embedding in each region is separately implemented using bin histogram.

A very popular method for reversible image watermarking is based on Prediction Error Expansion (PEE) [3]. When overflow/underflow happens, the pixels are left unchanged, and embedding is performed on one of the upper diagonal neighbors with different prediction errors.

In spite of many works on reversible watermarking, few types of research can be found on reversible video watermarking [4]. A video signal is composed of consecutive frames. One possible solution for reversible video watermarking is to watermark single frames with reversible watermarking algorithms developed for images. However, this solution is not efficient, since the temporal correlation between neighboring frames is not utilized. Reversible video watermarking methods with acceptable performance try to exploit interframe correlation. Fast motion video has a low temporal correlation, then watermarking with acceptable performance is another challenge.

Authors of [4], motion compensated interpolation error is used for reversible video watermarking. Unlike other methods that use motion vectors, [4] applies interpolation error for increasing information capacity. The capacity of each frame is adaptively specified, so distortion distribution is equalized among frames.

Liu *et al.* [5], have proposed a new robust reversible watermarking method for H.264 video. The watermark is embedded into Discrete Cosine Transform (DCT) coefficients of 4×4 blocks of frames. In [6] and [7], reversible watermarking for H.264 compressed video is proposed. In [6], histogram shifting of motion vectors is used. In [7], depth information is embedded based on quantized DCT coefficients expansion. Chung *et al.* [8] have proposed a reversible watermarking method based on prediction in mosaic videos. Chung has adaptively decreased the prediction error using either spatial or temporal correlations.

The PEE method is shown to provide better performance compared to other methods [9], i.e., higher capacity and lower distortion. In PEE method, the lower prediction error is equivalent to lower distortion and higher embedding capacity. Exploiting both temporal and spatial correlation in videos can help reduce prediction error, i.e., decrease distortion and increase the capacity [3] and [4].

Statistical properties of the image change from a region to another. Consequently, adaptive predictors can perform better than fixed predictors such as Median Edge Predictor, Gradient Adjusted Predictor, etc. [9].

In this paper, PEE method is used for reversible watermarking of medical videos. The proposed prediction scheme is linear and adaptive. Prediction coefficients are obtained by least square error method. To consider both temporal and spatial correlation, pixel $x_{i,j}$ in frame $k$ is predicted based on rhombus neighbors in $k^{th}$ frame, and its corresponding pixels in frames $(k+1)$ and $(k+2)$.

The rest of this paper is organized as follows. The proposed method is discussed in section 2. Performance of the proposed method is investigated in section 3. Finally, the contribution of our paper is summarized in section 4.

## II. PROPOSED METHOD

In part *A* of this section, we review PEE basics. Then in part *B*, we describe the proposed method.

### A. Overview of PEE Algorithms

In this section, we briefly explain PEE reversible image watermarking and linear prediction methods based on [3]. In embedding stage, given the pixel $x_{i,j}$ and its prediction $\hat{x}_{i,j}$, the prediction error is calculated by

$$e_{i,j} = x_{i,j} - \hat{x}_{i,j} \qquad (1)$$

Let us take a threshold $t$ to create a balance between the imperceptibility and the embedding capacity. If absolute prediction error is less than $t$, then the new value of the pixel is calculated by (2) and a bit of watermark, $b$, is embedded:

$$x'_{i,j} = x_{i,j} + e_{i,j} + b \qquad (2)$$

If the prediction error is greater than or equal to $t$, the pixel cannot be used for embedding and the pixel should be modified (shifted) by (3). In the extraction stage, this shift will show a greater prediction error, as compared to embedded pixels, and hence, it would guarantee reversibility.

$$x'_{i,j} = \begin{cases} x_{i,j} + t & if \quad e_{i,j} \geq t \\ x_{i,j} - (t-1) & if \quad e_{i,j} \leq -t \end{cases} \qquad (3)$$

In case of overflow/underflow, the pixels are left unchanged. Consequently, it is crucial for the extraction stage to differentiate between changed and unchanged pixels. If there are few unchanged pixels, one possible solution is to keep a list of their coordinates. If the number of overflow/underflow pixels is large, we may provide a map of overflow/underflow pixels (e.g., 0 for unchanged pixels and 1 for changed pixels). The coordinate list or pixels map is embedded into the frame in raw/compressed form for extraction stage.

Let us suppose that the extraction procedure should have the reverse order of the embedding procedure. In other words, the first extracted pixel should be the last embedded one. In the extraction stage, for a given pixel $x'_{i,j}$ and its prediction $\hat{x}_{i,j}$, the prediction error is calculated by:

$$e'_{i,j} = x'_{i,j} - \hat{x}_{i,j} \qquad (4)$$

Embedded pixels and shifted pixels are differentiated by prediction error. When overflow/underflow occurs we know by the mask. Otherwise, if the condition $(-2t + 2 \leq e'_{i,j} \leq 2t - 1)$ holds it is concluded that an embedded pixel is present. When neither of the mentioned conditions holds then the pixel is a shifted pixel. The original value of shifted pixels is reconstructed by the inverse of (3). For the embedded pixels, the original value is reconstructed by:

$$x_{i,j} = \frac{x'_{i,j} + \hat{x}_{i,j} - b}{2} \qquad (5)$$

where the watermark bit $b$, is extracted from the LSB of $e'_{i,j}$.

In linear prediction methods, pixel $x_{i,j}$ is predicted by a weighted sum of certain neighboring pixels via (6), as suggested in [9]:

$$\hat{x}_{i,j} = floor(\sum_{m=1}^{n} x_{i,j}^m v_m) \qquad (6)$$

where $x_{i,j}^1, \ldots, x_{i,j}^n$ are the prediction contexts including the spatial and temporal neighbor pixels, $n$ is the predictor order, $m$ is the index of prediction context, and $V = [v_1, \ldots, v_n]$ is the row vector of predictor coefficients. One possible solution for finding predictor coefficients $(V)$ is the least square error method, as shown below:

$$V = (X_{i,j}{}^T X_{i,j})^{-1} X_{i,j}{}^T y_{i,j} \qquad (7)$$

where $X_{i,j}$ is a matrix whose rows are the contexts vectors ($[x_{i,j}^m]$) in (6) and $y_{i,j} = [x_{i,j}^m]^T$ is the context prediction vector.

For extraction, we want to calculate the prediction error. For this purpose, we need to know the original pixel value, $x_{i,j}$. However, the original value are not available in extraction stage. Consequently, we have to estimate the pixel value by averaging their context pixel. As long as the pixels have the same predicted value of (6), reversibility of watermarking is obtained. Therefore, before the calculation of $X_{i,j}$ and $y_{i,j}$, pixel $x_{i,j}$ is replaced by the estimated value.

### B. Proposed watermarking algorithm

#### 1) Embedding procedure

A flowchart of the embedding phase is presented in Fig. 1. Inputs to the system are the video frames and the watermark logo. Size of the watermark we use is constant in all frames. The system outputs are the watermarked video and a 32-bit string named 'S', which contains required extraction information. Coordinates of the final embedded pixel take 19 bits of S, and 3 bits are used to indicate the value of the threshold and 1 bit is for the checkerboard set, as explained in the next section.

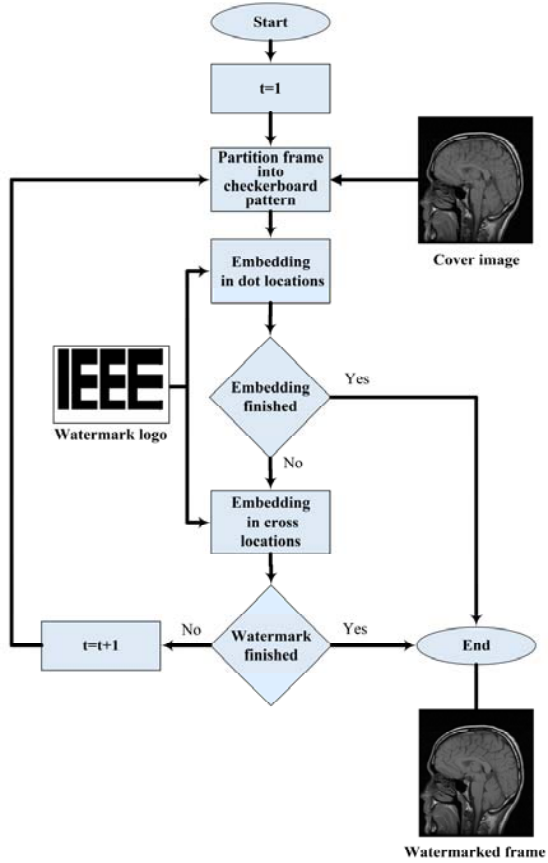

Figure 1: Flowchart of the embedding procedure.

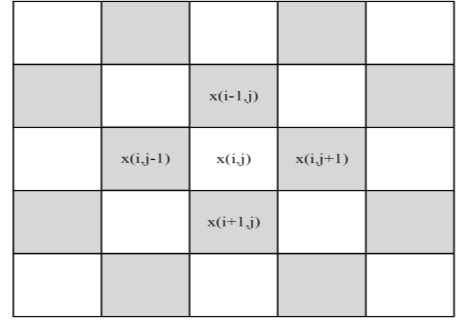

Figure 2: The chessboard pattern used for embedding order of pixels.

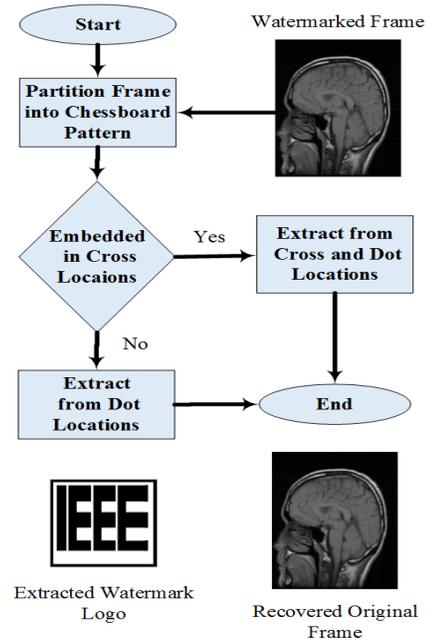

Figure 3: Flowchart of the extraction procedure.

We exploit the spatial correlation of neighboring pixels, to reduce the prediction error. The algorithm works on the diagonal pixels of a chessboard pattern, as shown in Fig. 2. Hence, pixels on the chessboard pattern are divided into two distinct sets of cross (black) and dot (white). Embedding starts from the dot set pixels, until all of them are used up. The cross pixels will be used after processing all the dot pixels, in case further unembedded information exists. With this approach, we avoid using the previous embedded pixels for prediction as much as possible. The embedding process continues until all the watermark bits are embedded. Row, column and set of last embedded pixels are stored in the string 'S'.

We also exploit temporal correlation by using the pixels in the next two neighbor frames, to reduce the prediction error. Hence, used frames for embedding are the unprocessed frames. The only exceptions are the last two frames in which their previous frames are used for prediction.

If any unembedded watermark bits remain after processing all pixels from both sets, we need to increase the threshold value for increasing the data capacity and start the whole process again. In the beginning, threshold $t$ is set to '1' and then increased by 1, if further capacity is required. This makes the system adaptive to the required capacity. Parameter $t$ is stored in the string 'S'.

Overflow/underflow pixels are left unchanged in the proposed method. We use a flag bit 'f' to differentiate between changed and unchanged pixels. These flag bits are embedded into the frame.

In the least square error method, according to the predictor order $n$, a number of border pixels cannot be watermarked. In our method, border pixels in two-pixel thickness, cannot be embedded.

As mentioned before, in the least square method, the original value of pixels $x_{i,j}$ cannot be used for their own prediction. Instead, we have to estimate a value for pixel. For this purpose, we use a simple rhombus predictor for estimating the pixel $x_{i,j}$ is predicted by fixed rhombus predictor (8). Since we want to consider temporal correlation, its corresponding pixel in next frame are considered in (8). Let superscripts be frame number.

$$\hat{p}_{i,j} = round(\frac{x_{i-1,j}^k + x_{i+1,j}^k + x_{i,j-1}^k + x_{i,j+1}^k + x_{i,j}^{k+1}}{5}) \quad (8)$$

For each pixel $x_{i,j}$, the embedding algorithm is:

**Embedding Algorithm**

1: Calculate $\hat{p}_{i,j}$ by equation (8) and create $X_{i,j}, y_{i,j}$

2: Calculate $V$ by equation (7) and $\hat{x}_{i,j}$ by equation (6)

3: Calculate $e_{i,j}$ by equation (1)

4: **If** $t < |e_{i,j}|$, **then** calculate $x'_{i,j}$ by equation (2) (embedding)

5: **Else** calculate $x'_{i,j}$ by (3) (shifting).

6: **If** $x'_{i,j} \in [0,255]$ (no overflow/underflow), **then**

7: Replace $x_{i,j}$ by $x'_{i,j}$

8: **If** $(x'_{i,j} \leq (t-2)$ or $x'_{i,j} \geq (256-t))$, **then** insert flag bit f=1 into the next embeddable pixel

9: **Else** insert flag bit f=0 into the next embeddable pixel

*2) Extraction procedure*

Flowchart of the extraction procedure is presented in Fig. 3. Inputs are watermarked video frames and the string 'S'. Outputs are reconstructed video frames and the watermark logo.

For each pixel $x'_{i,j}$, the extraction algorithm is:

**Extraction Algorithm**

1: Calculate $\hat{p}_{i,j}$ by (8) and create $X_{i,j}, y_{i,j}$

2: Calculate $V$ by (7) and $\hat{x}_{i,j}$ by (6)

3: Calculate $e'_{i,j}$ by (4)

4: **If** $(-2t+2 \leq e'_{i,j} \leq 2t-1)$, **then**

5: **If** $(x'_{i,j} \leq (t-2)$ or $x'_{i,j} \geq (256-t))$, **then**

6: **If** its corresponding flag bit is '1', **then**

7: Extract the watermark bit, $b$, from the LSB of $e'_{i,j}$ and reconstruct the original value by (5)

8: **Else** an unchanged pixel is reported

9: **Else** extract the watermark bit, $b$, from the LSB of $e'_{i,j}$ and reconstruct the original value by (5)

11: **Else**

12: **If** $(x'_{i,j} \leq (t-2)$ or $x'_{i,j} \geq (256-t))$, **then**

13: **If** its corresponding flag bit is '1', **then**

14: Reconstruct original value by inverting (3).

15: **Else** an unchanged pixel is reported.

16: **Else** reconstruct original value by inverting (3).

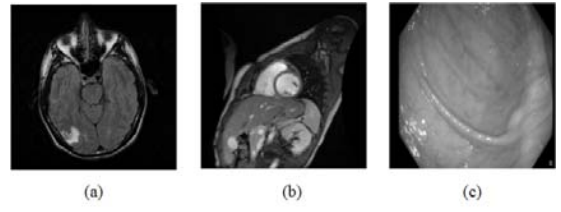

Figure 4: (a) Brain MRI, (b) Cardiac MRI, (c) Intestinal Polyp.

### III. RESULTS

Performance of the described reversible watermarking algorithm is tested on three medical video sequences (Brain MRI [11], Cardiac MRI [12] and Intestinal Polyp [13]). Test videos demonstrate different motion rates. The Cardiac MRI sequence has moderate motion, the Brain MRI sequence has fast motion, and Polyp sequence has slow motion. Sample frames of the three datasets are shown in Fig. 4 for visual demonstration. Each sequence includes 18 frames. The frame size of Brain MRI and Cardiac MRI sequences is 256×256, but Polyp frames have a bigger size of 480×856. For the input watermark, we use a binary image with an equal distribution of ones and zeros (49% ones and 51% zeros).

In Fig. 5 original frame and watermarked frame for each sequence is shown. It can be seen that watermarked frame is visually identical to the original frame. For evaluating the performance of our algorithm, we calculate the capacity and distortion of the watermarked frames. Watermarking capacity in each frame is calculated by the ratio of a total number of watermark bits to the total number of pixels in the frame, expressed in bit per pixel (BPP):

$$BPP = \frac{L}{W \times H} \quad (9)$$

where $L$ is the total number of watermark bits, $W$ and $H$ are the frame's width and height respectively.

For evaluating the distortion of frames, we have used Peak Signal to Noise Ratio (PSNR) [10] as a metric, which is described in (10):

$$PSNR = 10 \, log_{10} \frac{255^2}{\frac{1}{N \times W \times H} \sum_{k,i,j=1}^{N,W,H} (F_{i,j}^w(k) - F_{i,j}(k))^2} \quad (10)$$

where $N$ is count of frames, $W$ and $H$ are width and height of frames respectively. $F_{i,j}(k)$ and $F_{i,j}^w(k)$ are pixel $(i,j)$ in frame $k$ of original and watermarked videos respectively.

Table 1 summarizes the capacity-distortion results of our proposed method for the three medical sequences. We divide the video sequence into 3 groups, where each group of frames contain 6 frames. Table 1 indicates that distortions in different groups of one sequence are almost equal and even near to that of the whole sequence which is the union of the groups in that sequence. These results also demonstrate that the proposed method is almost independent of the number of the frames of sequence. As we can see watermark image of size 52×65 is successfully embedded into the fast motion MRI sequence.

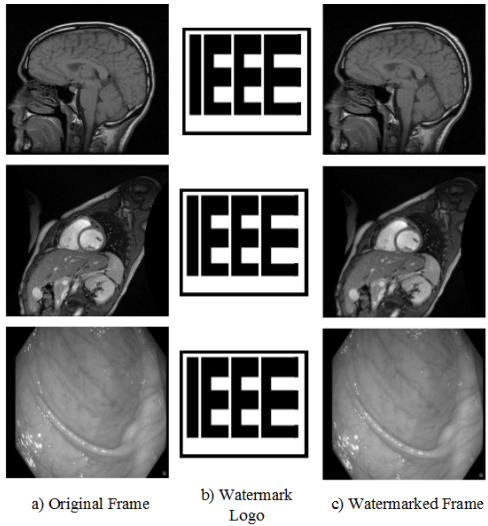

a) Original Frame  b) Watermark Logo  c) Watermarked Frame

Figure 5: The visual quality of watermarked frames using the proposed embedding method.

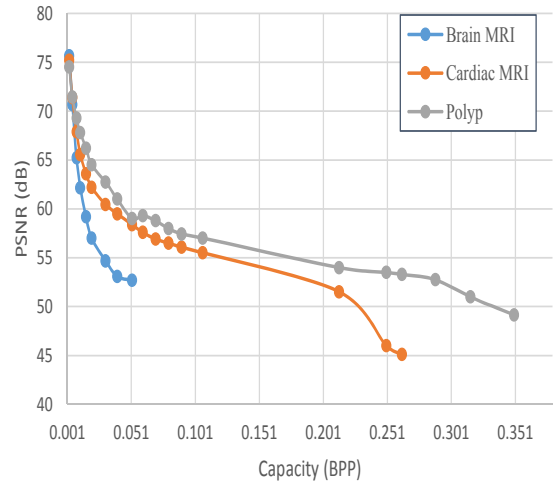

Figure 6: PSNR as a function of capacity for the proposed method.

TABLE I. AVERAGE PSNR FOR DIFFERENT CAPACITIES.

| Watermark size | | *32×32* | *45×45* | *52×65* |
|---|---|---|---|---|
| Average PSNR of Brain MRI video (dB) | *Group 1* | 58.8585 | 54.6089 | 52.6738 |
| | *Group 2* | 59.9209 | 54.9546 | 53.1324 |
| | *Group 3* | 58.5134 | 54.5059 | 52.286 |
| | *All groups* | 59.187 | 54.6817 | 52.6942 |
| Average PSNR of Cardiac MRI Video (dB) | *Group 1* | 63.6774 | 60.4947 | 58.3929 |
| | *Group 2* | 63.4404 | 60.3482 | 58.3454 |
| | *Group 3* | 63.4382 | 60.3423 | 58.3062 |
| | *All groups* | 63.5875 | 60.4381 | 58.3764 |
| Average PSNR of Intestinal Polyp video (dB) | *Group 1* | 74.1876 | 71.0835 | 68.7212 |
| | *Group 2* | 74.0803 | 71.1009 | 68.7336 |
| | *Group 3* | 74.8347 | 71.6936 | 69.3134 |
| | *All groups* | 74.5175 | 71.4508 | 69.0291 |

When Embedding the small watermark of size 32×32 in Polyp sequence, results in PSNR=74dB, implying a very low distortion. It is shown that increasing the watermark size to almost double, doesn't lead to proportional increase of distortion, i.e. for the watermark size of 52×65. The proposed method provides an acceptable low distortion on all of the three medical sequences of our experiments.

Fig. 6 shows distortion-capacity results of the three experiments. Three observations may be noticed in these plots. Firstly, increasing the watermark size, does not lead to a proportional increase in distortion. Secondly, the faster is the motion; a lower capacity is available for reversible watermarking. For the fast motion Brain MRI, maximum capacity is 0.051 BPP, while the slow Polyp sequence provides a maximum capacity of 0.351 BPP. Thirdly, the faster is the motion; we observe a higher distortion. The Cardiac MRI curve, demonstrate 46 dB PSNR for a capacity of 0.251 BPP, while the Polyp curve shows 54 dB PSNR for the same embedding capacity.

IV. CONCLUSION

In this paper, we have investigated a new method for reversible watermarking in medical videos. This means that original videos can be reconstructed from the watermarked frames. Unlike previous methods, which embedding and extraction of watermarks are based on motion vectors, we apply a linear adaptive predictor. To increase the watermark capacity, the temporal correlation has been exploited for prediction. We have shown that watermark image of size 52×65 can be embedded in the fast motion Brain MRI with very low distortion (PSNR=52dB). The proposed method is also applicable for non-medical videos.